%% file: 2130.tex
\documentclass{aa}
\usepackage{pictexwd,rotating,amssymb,nicefrac,verbatim,graphics}
\usepackage{booktabs,afterpage}
\begin{document}

   \title{A QSO survey via optical variability and zero proper motion
   in the M\,92 field}
   \subtitle{III. Narrow emission line galaxies}

   \author{H. Meusinger\,$^{\star}$
          \and
           J. Brunzendorf
\thanks{
Visiting Astronomer, German-Spanish Astronomical Centre, Calar Alto,
operated by the Max-Planck-Institute for Astronomy, Heidelberg, jointly
with the Spanish National Commission for Astronomy}
          }

   \institute{Th\"uringer Landessternwarte Tautenburg, D-07778 Tautenburg,
   Germany\\
              e-mail: meus@tls-tautenburg.de, brunz@tls-tautenburg.de
             }


   \date{Received 26 November 2001 / Accepted 13 May 2002}

   \abstract{
   We study a sample of 23 narrow-emission line galaxies (NELGs)
   which were selected by their strong variability
   as QSO candidates in the framework of a
   variability-and-proper motion QSO survey on digitised Schmidt plates.
   In previous work, we have shown that variability is an
   efficient method to find AGNs. The variability properties of
   the NELGs are however significantly different from those
   of the QSOs. The main aim of this paper is to clarify
   the nature of this variability and to estimate the fraction of
   AGN-dominated NELGs in this sample.
   New photometric and spectroscopic observations
   are presented, along with revised data from the
   photographic photometry. The originally measured high variability
   indices could not be confirmed. The diagnostic line-ratios of 
   the NELG spectra are consistent with \ion{H}{ii} region-like spectra.
   No AGN could be proved, yet we cannot rule out the existence of
   faint low-luminosity AGNs masked by \ion{H}{ii} regions from
   intense star formation.
   }

\authorrunning{Meusinger \& Brunzendorf}
\titlerunning{VPM QSO survey in the M\,92 field.III. NELGs}

   \maketitle

%
\section{Introduction}

The variability of flux densities is a common property of
high-luminosity AGNs. We have performed a QSO search based on variability and
proper motion (VPM survey) measured on a large number of digitised
Schmidt plates in two fields (Meusinger et al. \cite{Meu02}).
The work in the M\,92 field is the subject of the present series of papers.
In the first paper (Brunzendorf \& Meusinger \cite{Bru01}; hereafter
Paper\,1), we discussed the motivation, the observational
data, the data reduction procedure, and the selection of QSO
candidates. The results from the follow-up spectroscopy
and the properties of the resulting QSO sample were the
subject of Paper\,2 (Meusinger \& Brunzendorf \cite{Meu01}).
The primary goal of the present study is to improve the
understanding of the selection effects of this survey.

An object is considered a VPM-QSO candidate if it appears star-like,
has no significant proper motion, and shows significant overall variability
and long-term variability. The variability  is
expressed by the indices $I_\sigma$ (overall variability) and $I_\Delta$
(long-term variability). For instance, an object with $I_\sigma>2$ has
a probability of $\alpha>0.98$ to be variable.
It is well known that high-luminosity AGNs vary on long timescales
(years and longer). High priority QSO candidates have therefore to meet
both $I_\sigma\ge2$ and $I_\Delta\ge2$.
On the other hand, we found several QSOs with strong overall variability
but without significant long-term variability (Paper\,2). The long-term
variability constraint may introduce a bias in the VPM QSO search, and
it is therefore important to study also the subsample of
variable, star-like objects with zero proper motion showing
no significant long-term variability. In particular, we found
27 narrow emission line galaxies (NELGs) with redshifts $z \la 0.2$
in this subsample. Most of these galaxies show
strong emission lines. NELGs may be dominated by
narrow-emission line AGNs (Seyfert\,2, narrow-emission line Seyfert\,1,
LINERs), intense starbursts, or a mixture of both. For example,
Ho et al. (\cite{Ho97}) found that about half of
the NELGs from their magnitude-limited sample show some form of AGN or
composite spectra. In Paper~2, we have speculated that at least some
of the VPM NELGs are dominated by AGNs, though the available data did
not allow a clear-cut conclusion.

The present paper is concerned with the sample of the NELGs
from the VPM survey.  The main question is whether the measured strong
overall variability as well as the strong emission lines are
related to AGNs or not. It is not our intention to provide
a large and well-defined sample of NELGs useful for further
detailed studies. Much larger samples (e.\,g.,
Terlevich et al. \cite{Ter91};
Ho et al. \cite{Ho97};
Popescu \& Hopp \cite{Pop00})
are available and are better suited to the investigation of
the overall NELG population. In Sect.\,2, we present
new spectroscopic and photometric observations. Section\,3 is concerned
with the variability properties of the NELGs. The spectroscopic
properties are discussed in Sect.\,4, and Sect.\,5 reviews
further properties of the galaxy sample. Sect. 6 concludes.
As in the previous papers of this series,
we adopt $H_0 = 50$\,km\,s$^{-1}$\,Mpc$^{-1}$ and $q_0=0$.

%
\section{Observations and data reduction}

\input 2130f01

Low resolution spectra of VPM QSO candidates were described in
Paper~2. Unfortunately, the spectra for the NELGs did not 
allow a clear-cut separation between the principal ionisation sources
(AGNs versus massive stars). New
observations were performed with CAFOS at the 2.2\,m telescope
on Calar Alto, Spain, during six nights in July 2000.
CAFOS was equipped with a SITe1d CCD. The grism was chosen
dependent on the redshift: low-$z$ NELGs were observed
with G-100 in order to achieve a good separation between
H$\alpha$ and [\ion{N}{ii}]$\lambda$6583\,\AA{}. For the
higher-$z$ NELGs, G-200 was used because of its higher
transmission at longer wavelengths. For some objects, spectra were taken
with both grisms. Total integration times between 30 and 60\,min were
necessary to obtain spectra of reasonable signal-to-noise.
The weather conditions were mainly good.
The seeing was stable ($1\farcs0$ to $1\farcs2$) and the slit width was
kept constant, resulting in a linear resolution of 10\,\AA{} (G-200)
and 5\,\AA{} (G-100), respectively. The orientation of the slit was always
North-South. Wavelength calibration spectra were taken
by means of Hg-He-Ar calibration lamps.

We omitted two of the 27 NELGs that are
located close to brighter galaxies since their measured variability is very
likely not real. Further, the two NELGs of lowest priority could not be
observed due to poor weather on the last night of the observing run.
For the remaining 23 NELGs, spectra of good quality were obtained. In addition,
six comparison galaxies with well-known spectroscopic data and
spectral classification (see Table~3) were observed.
All spectra were reduced on the basis of ESO-MIDAS routines, in
particular the MIDAS package LONG.  The resulting one-dimensional
spectra are dominated by the emission from the central
regions of the NELGs. The spectra were not flux-calibrated.

A subset of NELGs were photometrically monitored during the CAFOS campaign.
We selected the 10 galaxies with highest variability indices and small
deviations from a star-like image structure. On each of
the six nights, we took a 180\,s direct image of a
$5\arcmin\times5\arcmin$ field around each of the 10 galaxies through a
Johnson $B$ filter.  The CCD frames were reduced using MIDAS
standard routines.

%
\section{Variability}

\input 2130f02

The variability indices derived from the magnitude measurements
in Paper\,1 show two fundamental differences between
NELGs and QSOs. First, there is no indication for
significant long-term variability in the NELG sample, contrary to
the QSOs. On the other hand, strong variability of the NELGs
is indicated at short timescales of a few days or less.
These differences are clearly
illustrated by the comparison of the sample-averaged
structure functions in Fig.\,\ref{sf} (for definitions and
properties of the structure function see e.g. Simonetti et al.
\cite{Sim85}). The absence of long-term variability is probably
consistent with the presence of
low-luminosity AGNs (LLAGNs) that are suspected to have
shorter variability timescales than high luminosity AGNs
(Filippenko \cite{Fil92}; Lira et al. \cite{Lir99};
Moran et al. \cite{Mor99}).

Strong variability on short timescales is provable
by means of CCD time-series observations with a baseline of a few days.
The results from our CCD-monitoring campaign do not indicate
significant variability at the 0.02\,mag level.
We adopt the null-hypothesis
$H0:\ \sigma_{B,{\rm NELG}} = \sigma_{B,{\rm star}}$,
i.\,e. the photometric standard deviation $\sigma_B$ is the same
for the NELGs and the stars of comparable magnitude. The $F$ test
shows that $H0$  should not be rejected on  a significance level
$\alpha = 0.95$ for 9 of the 10 NELGs. For the remaining object,
the measured $F$ is close to the critical value $F_{\rm crit}$.
Hence, there is no evidence of significant short-term variability
from the CCD time-series.

The photographic photometry described in Paper\,1 was
based on a two-dimensional
Gaussian profile-fitting procedure. Deviations of the image profile from
the Gaussian leads to an increased measurement error and, combined with
variations of the observing conditions from plate to plate, to
artificial variability. This effect is clearly reflected by  the high
variability indices measured for galaxies with extended images. However,
all but three NELGs appear star-like even on the deepest plates, and
none of the NELGs was classified as extended on the basis of the
index $I_{\rm nonstellar}$.

In order to solve the discrepancy between the strong variability
from the Schmidt plate data and the results of the CCD time-series,
we have completely revised the reduction of all 162 $B$ Schmidt
plates. The SExtractor package (Bertin \& Arnouts \cite{Ber96})
was used and magnitudes were measured by aperture photometry instead of
profile fitting. At the faint end, the photometric accuracy of the revised
data is improved by a factor of about two. An additional
improvement is achieved by averaging the measured magnitudes
of an object over adjacent epochs. This was done if
(a) the single-epoch data have too low a photometric accuracy, i.e.
$\sigma_B(B=19^{\rm m})>0.1$ or $\sigma_B(B=20^{\rm m})>0.2$, and
(b) there are several plates available of close-by epochs.
Thus, we finally have 54 data points of different epochs spanning 33.2 years.
For each individual epoch, the photometric accuracy is better than
0.1\,mag at $B=19$ and better than 0.2\,mag at $B=20$. This is a significant
improvement compared to the original data (cf. Fig.~5 in Paper~1).

\begin{figure}[hpbt]
\resizebox{9.0cm}{7.2cm}{\includegraphics{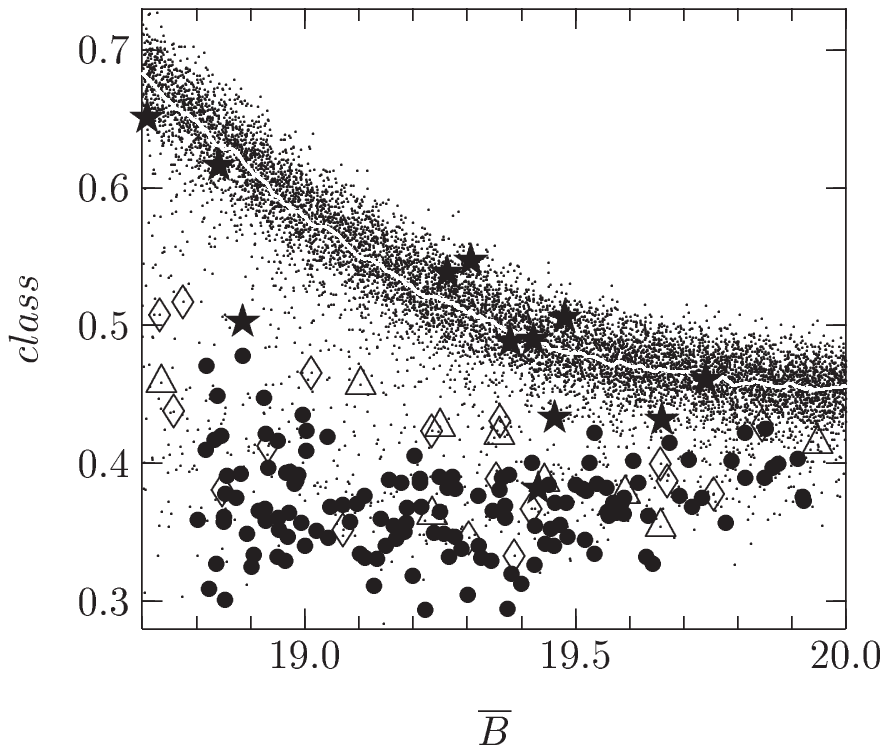}}
\caption{\label{class}
The new object classification parameter {\it class} for all objects 
(dots) in the magnitude range of the NELGs. 
The white curve represents the running median,
the various symbols designate visually identified galaxies (bullets), 
Seyfert\,1 galaxies with $z \le 0.5$ (asterisks), and NELGs 
(open lozenges for $M_B<-20$, open triangles for $M_B>-20$),
respectively.  
}
\end{figure}

\begin{table*}
\caption{\label{foto}
Photometric data for the 23 NELG sample of the present study.
The running number, J2000.0 position, mean $B$ magnitude,
mean absolute magnitude $M_B$ (with redshift from Table~\ref{spectra}),
old and new overall variability indices $I_\sigma$ and $I_\Delta$,
new morphology index $I_{\rm nst}$ (where ``nst'' means ``nonstellar''),
as well as the mean colour indices $U-B$ and $B-V$ (without reddening
correction, without $k$ correction) are given.
}
\begin{tabular}{rlccrrrrrrr}
\toprule
no. & NELG          &$\overline{B}$ & $M_B$
& $I_\sigma^{\rm old}$ & $I_\Delta^{\rm old}$
& $I_\sigma^{\rm new}$ & $I_\Delta^{\rm new}$
& $I_{\rm nst}^{\rm new}$
& $\overline{\mbox{\it U--B}}$ & $\overline{\mbox{\it B--V}}$\\
\midrule

 1 & J171122.0+440721 & 19.4 &  -20.5  &   5.53  &  0.33 &  1.44  &  0.69 &   3.7 & -0.45 & 0.80 \\ 
 2 & J171124.1+433117 & 19.7 &  -21.0  &   5.19  & -0.57 &  2.87  & -1.38 &   5.1 & -1.38 & 0.29 \\ 
 3 & J171241.1+430512 & 19.4 &  -21.0  &  13.66  & -2.00 &  3.11  &  0.11 &   6.7 & -0.17 & 1.10 \\ 
 4 & J171319.5+435216 & 19.1 &  -20.6  &  11.68  &  1.54 &  4.20  & -0.14 &  10.8 & -0.90 & 0.38 \\ 
 5 & J171323.0+431230 & 18.7 &  -20.6  &   9.19  &  1.75 & -0.72  & -0.29 &   8.7 & -0.33 & 0.65 \\ 
 6 & J171326.8+440117 & 19.1 &  -17.1  &   6.16  &  0.91 & -0.03  & -0.22 &   4.7 & -0.52 & 0.76 \\ 
 7 & J171448.3+434455 & 18.8 &  -20.7  &   8.96  &  0.37 &  0.69  &  1.66 &   7.4 & -0.24 & 0.43 \\ 
 8 & J171459.0+434327 & 19.4 &  -18.4  &   7.57  &  1.38 &  0.07  &  1.21 &   5.6 & -0.07 & 0.85 \\ 
 9 & J171510.7+430506 & 19.4 &  -20.1  &   6.41  &  1.30 &  0.74  &  0.17 &   8.9 & -0.63 & 1.15 \\ 
10 & J171520.1+433427 & 18.7 &  -19.5  &  10.61  & -0.22 &  3.32  & -1.75 &  11.3 & -0.25 & 0.66 \\ 
11 & J171610.9+422333 & 18.8 &  -21.6  &  32.47  &  0.45 &  5.70  & -2.05 &  12.0 & -0.50 & 1.06 \\ 
12 & J171652.4+433528 & 19.7 &  -18.1  &   6.26  &  2.04 &  2.11  & -0.21 &   7.2 & -0.12 & 0.51 \\ 
13 & J171734.2+432824 & 19.9 &  -18.8  &   7.92  &  1.31 &  0.87  & -0.04 &   2.9 & -0.79 & 0.67 \\ 
14 & J171734.9+425643 & 19.8 &  -20.3  &   6.78  & -0.18 & -1.00  &  0.26 &   2.0 & -0.62 & 0.73 \\ 
15 & J171747.3+432550 & 19.2 &  -18.8  &  10.82  &  0.44 &  2.85  & -0.86 &   8.2 & -0.52 & 1.05 \\ 
16 & J171828.1+442727 & 19.4 &  -18.8  &   5.82  & -1.49 &  1.89  &  1.71 &   4.2 & -0.44 & 0.65 \\ 
17 & J171908.9+423111 & 19.3 &  -21.0  &  10.37  &  1.39 &  2.97  &  0.77 &   8.8 & -0.76 & 1.06 \\ 
18 & J171955.6+442244 & 18.9 &  -20.4  &  11.85  & -1.25 &  4.23  &  0.41 &   9.7 &  0.34 & 0.62 \\ 
19 & J172156.0+441912 & 19.3 &  -17.0  &   6.74  & -1.79 & -2.02  & -0.08 &   4.8 & -0.71 & 0.53 \\ 
20 & J172256.1+425447 & 19.4 &  -21.2  &   9.76  &  0.51 &  2.28  &  2.57 &   6.0 & -0.64 & 1.17 \\ 
21 & J172340.6+434102 & 19.6 &  -19.4  &  14.51  & -2.41 &  2.66  & -0.20 &   5.8 & -0.19 & 0.56 \\ 
22 & J172348.1+432907 & 19.0 &  -21.4  &   9.54  & -0.40 &  2.86  & -1.21 &   5.5 & -1.01 & 0.93 \\ 
23 & J172407.6+424037 & 19.2 &  -21.3  &   3.65  &  0.55 &  1.46  &  0.65 &   5.0 & -0.75 & 0.69 \\ 
\bottomrule
\end{tabular}
\end{table*}

The variability indices $I_\sigma$ and $I_\Delta$ were computed in
exactly the same way as in Paper~1. In Fig.~\ref{old_new},
we compare the new with the old variability indices of the NELGs.
The values are listed in Table\,\ref{foto},
along with the other photometric data.
As for the original data, no significant long-term variability
is found from the revised data. The overall variability indices
from the revised photometry are considerably reduced: only about
50\% of the NELGs have $I_{\sigma}^{\rm new} > 2$.
An outstandingly high $I_\sigma^{\rm new} \approx 35$ is found
for an NELG that is projected onto an extended foreground galaxy.
These facts illustrate that the way of measuring magnitudes
is of major importance for the assessment of variability.

In Paper~1, image profile indices were derived from the
radius-magnitude relation. The SExtractor package allows a more
sophisticated morphological classification based on a trained neural
network. According to the classification parameter $class$ derived by
SExtractor, NELGs are clearly separated from star-like objects
(Fig.~\ref{class}). On the other hand, the classification parameters of some
of the Seyfert galaxies from the VPM survey are similar to NELGs.
Analogously with the nonstellar index from Paper~1,
we define a new nonstellar index
\begin{displaymath}
I_{\rm nonstellar}^{\rm new} =
\frac{class - \overline{class(B)}}{\sigma_{class}(B)},
\end{displaymath}
where  $\overline{class(B)}$ is the
median of the classification  parameter of all objects at magnitude $B$,
and $\sigma_{class}(B)$ is the standard deviation.
Figure\,\ref{norm_diff} clearly shows that
the objects with higher (new) nonstellar indices
tend to have higher (new) variability  indices. This correlation
strongly suggests that
the measured large variability indices of the NELGs  are dominated
by measurement errors due to the deviations from
stellar appearance.

\input 2130f04

%
\section{Spectral classification}

The spectra cover the wavelength interval between about 4\,800\,{\AA}
and 8\,200\,{\AA} in the observer frame. Prominent lines are H$\beta$,
[\ion{O}{iii}]$\lambda$5007,
[\ion{O}{i}]$\lambda$6300,
H$\alpha$,
[\ion{N}{ii}]$\lambda$6583,
and the [\ion{S}{ii}]$\lambda\lambda$6717,6731 blend.
The accuracy of the emission line equivalent widths (EWs) is
determined primarily by the uncertainty in the level of the continuum.
The problem of a subjective bias was partly overcome by a
non-interactive measurement procedure. Briefly,
the level of the local continuum is estimated by means of median
and low-pass filters. Then, the EW is estimated at the position calculated
from the previously measured redshift. The redshifts and the
resulting EWs are listed in Table\,\ref{spectra}.
The errors given there are the random
errors derived by the measuring procedure.
In general, the Balmer lines will be blended.
We follow the approach by Popescu \& Hopp (\cite{Pop00})
to correct for the absorption by the underlying older stellar population:
for all galaxies with strong continuum emission ($EW({\rm H}\beta)<20$\,\AA{})
the measured $EW({\rm H}\beta)$ is increased by an assumed constant
absorption EW. A constant value of
$EW_{\rm abs}({\rm H}\beta) \approx 2$\,\AA{}
was found by McCall et al. (\cite{McC85}) and is in good agreement with
the data given by Ho et al. (\cite{Ho97}). H$\alpha$ absorption was corrected
for in the same way with
$EW_{\rm abs}({\rm H}\alpha) \approx 1.7$\,\AA{}
derived from the Ho et al. sample.

\begin{figure*}[hbpt]
 \begin{tabbing}
 \resizebox{8.8cm}{7.9cm}{\includegraphics{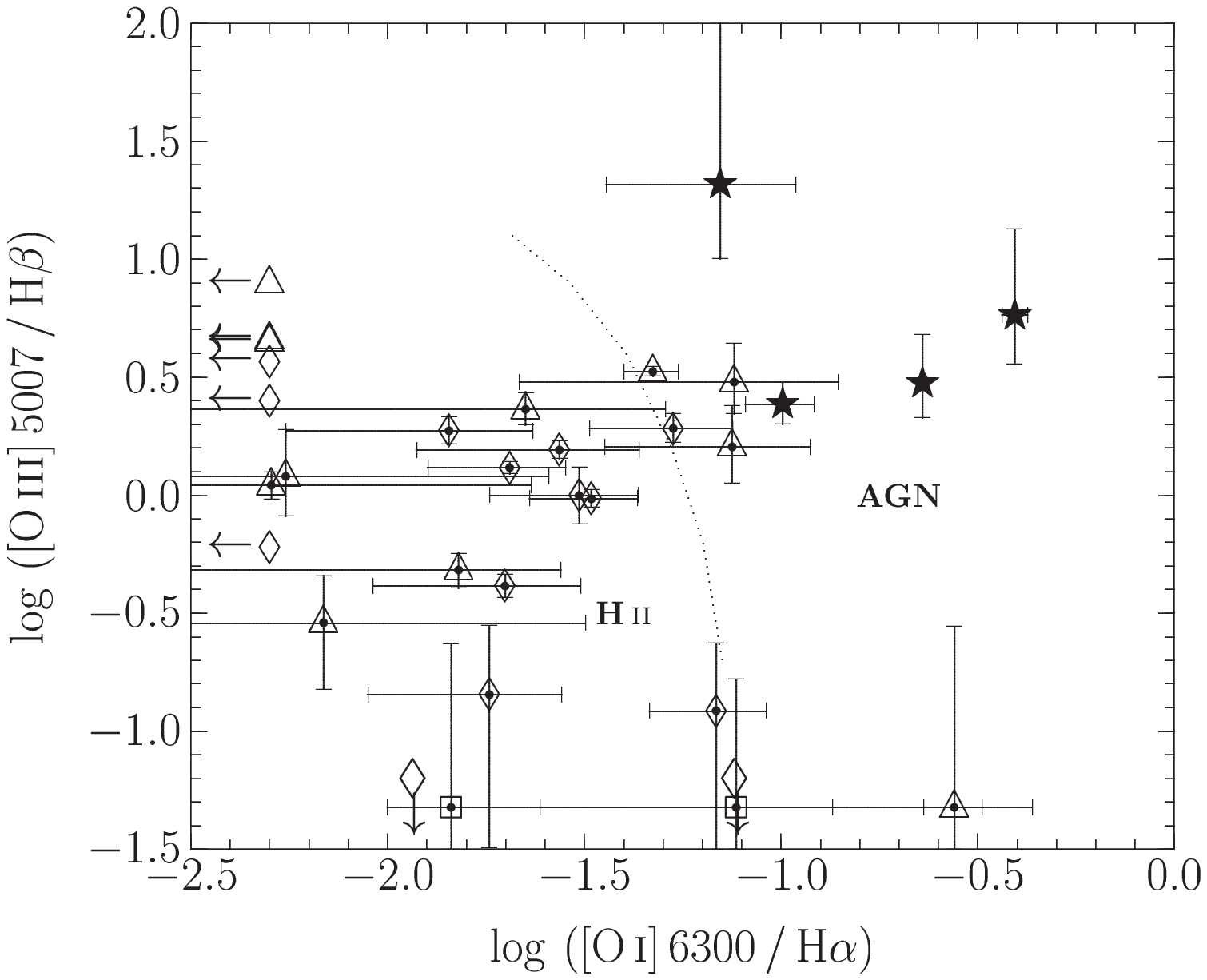}} \hfill \=
 \resizebox{8.8cm}{7.9cm}{\includegraphics{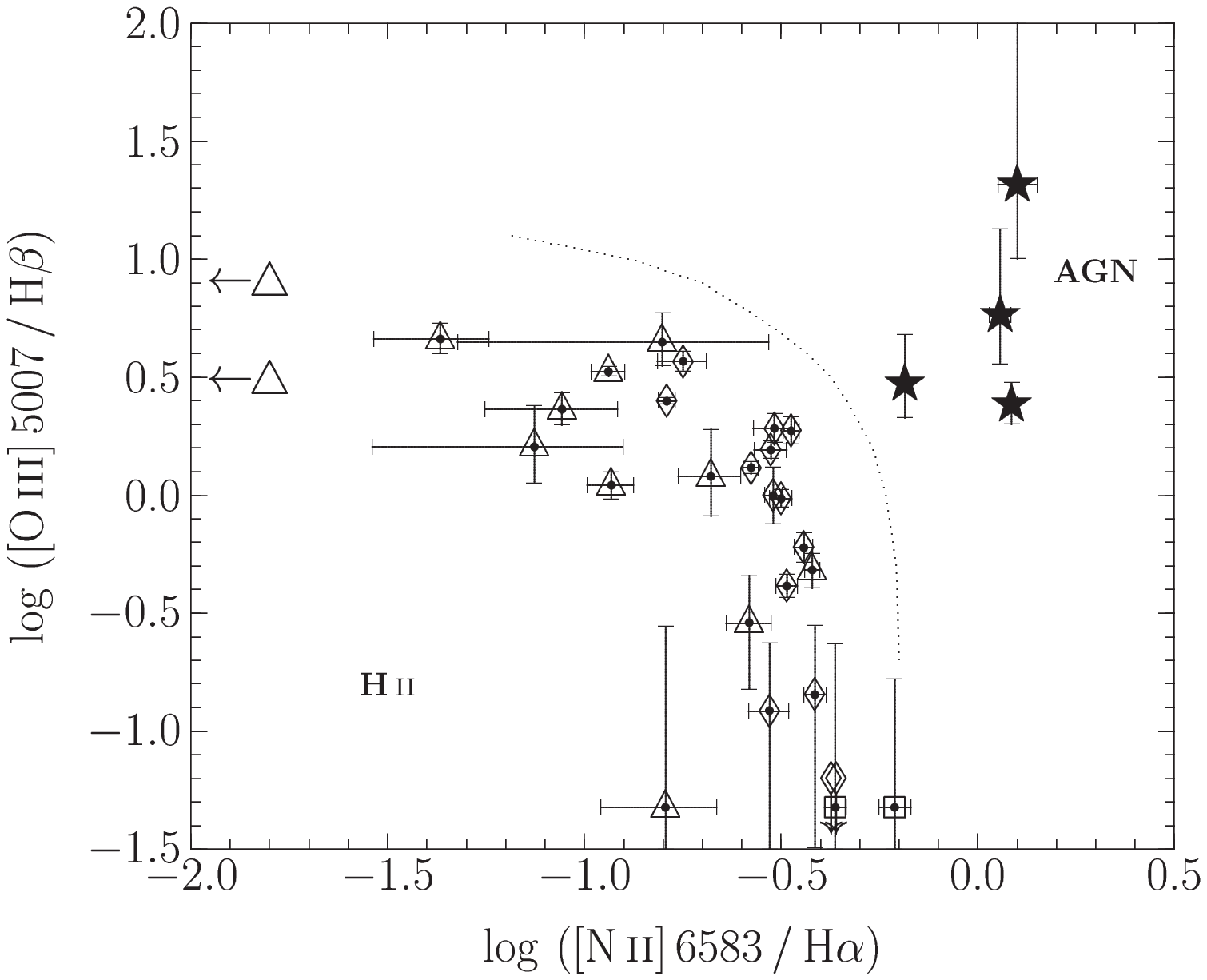}}\= \\
 \resizebox{8.8cm}{7.9cm}{\includegraphics{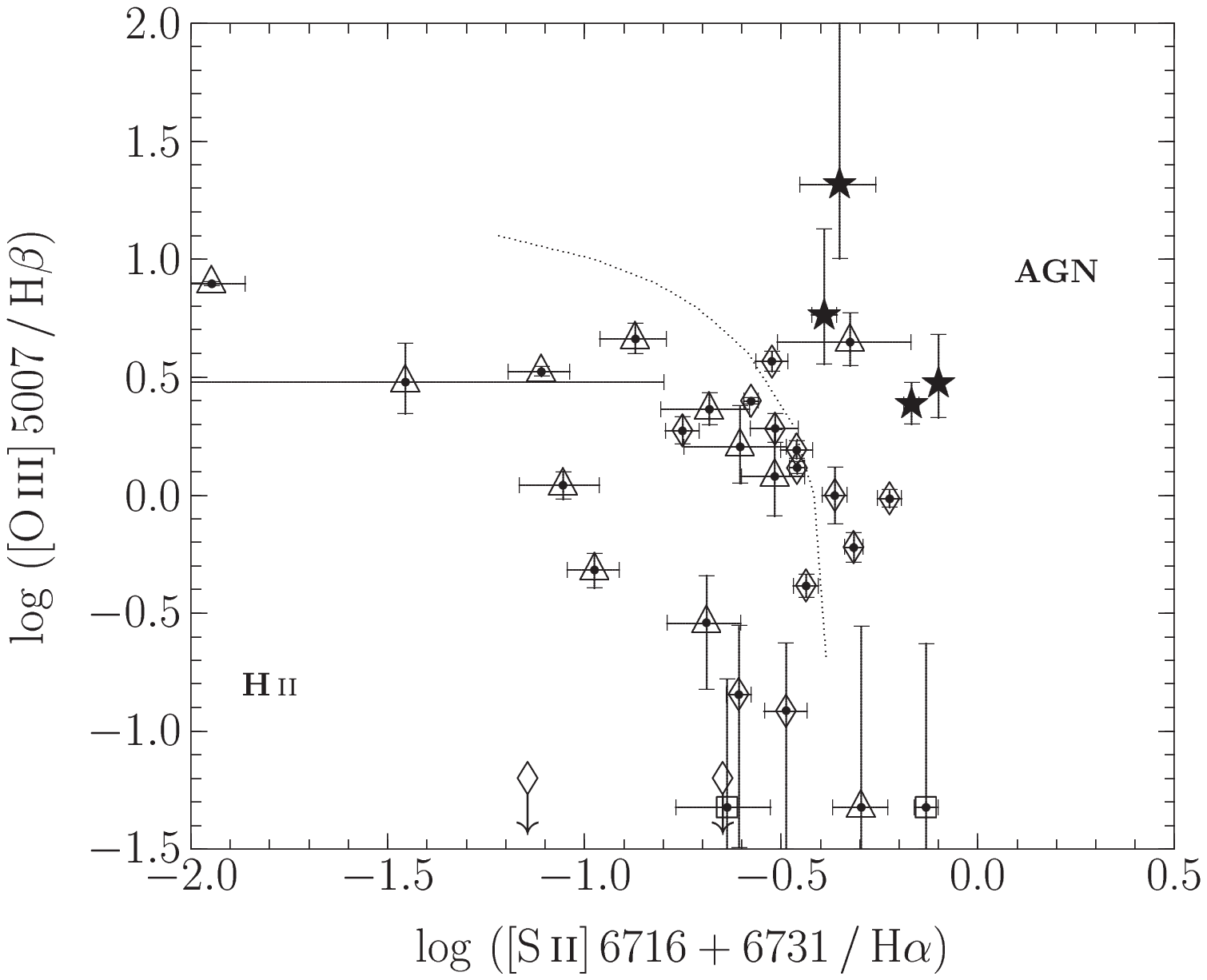}}\> \\
 \end{tabbing}
\caption{Diagnostic line-ratio diagrams for the NELGs (symbols as in
Fig.\,\ref{class}) and the comparison galaxies (galaxies classified in the literature
as Seyferts or LINERs are marked as asterisks, transition types
as open boxes).
\label{diagnostics}}
\end{figure*}

\begin{table*}[bpht]
\caption{\label{spectra}
Spectroscopic data for the NELG sample of the present study.
Running numbers from Table\,\ref{foto}, used grism, redshift $z$, 
measured equivalent widths (in \AA{}) of the major emission lines, 
and the results of the spectral classification are given. 
For ambiguous classifications the two
most likely classes are given with the most probable one at the first position. Classifications from spectra with
relatively low signal-to-noise ratio have a questionmark.
}
\begin{tabular}{rrcrrrrrrl}
\toprule
no. & grism &\quad\,$z$ &  H$\,\beta$\,\,\,\,  & [\ion{O}{iii}]\,\, & [\ion{O}{i}]\,\,\,\, & H$\,\alpha$\,\,\,\,\, & [\ion{N}{ii}]\,\,\,\,
& [\ion{S}{ii}]\quad\,\,\,\, & spectral\\
    &       &           &  4861\,\AA{} & 5007\,\AA{}    & 6300\,\AA{}  & 6565\,\AA{} &  6584\,\AA{}  & 6716/31\,\AA{} & class\\
\midrule
 1 & G100 &  0.1457 &  14.2$\pm$0.6 &   6.7$\pm$0.5  &  1.5$\pm$0.8 &  75.4$\pm$1.0 &  24.7$\pm$1.2 &  27.6$\pm$1.7 & \ion{H}{ii}     \\
 2 & G200 &  0.2087 &  30.7$\pm$0.6 &  77.0$\pm$0.7  &  0.1$\pm$1.2 & 171.5$\pm$1.1 &  27.8$\pm$1.2 &  45.5$\pm$1.7 & \ion{H}{ii}     \\
 3 & G200 &  0.1867 &   2.1$\pm$0.3 &   0.5$\pm$0.4  &  2.3$\pm$0.7 &  33.6$\pm$0.9 &   9.9$\pm$0.9 &  10.9$\pm$1.1 & \ion{H}{ii}     \\
 4 & G200 &  0.1362 &  22.3$\pm$0.8 &  74.5$\pm$0.9  &  6.1$\pm$0.9 & 129.1$\pm$1.3 &  14.9$\pm$1.3 &  10.0$\pm$1.7 & \ion{H}{ii}     \\
 5 & G100 &  0.1150 &   4.1$\pm$0.6 &  11.7$\pm$0.5  &  1.9$\pm$0.7 &  35.7$\pm$1.1 &  10.8$\pm$1.0 &  10.9$\pm$1.2 & \ion{H}{ii}     \\  
 6 & G100 &  0.0270 &   5.5$\pm$2.2 &   9.0$\pm$1.1  &  0.2$\pm$0.7 &  36.3$\pm$1.2 &   7.6$\pm$1.1 &  11.1$\pm$1.7 & \ion{H}{ii}     \\
 6 & G200 &  0.0275 &   7.9$\pm$0.7 &  10.9$\pm$0.7  &  0.2$\pm$0.7 &  39.5$\pm$0.6 &   4.6$\pm$0.6 &   3.5$\pm$0.8 & \ion{H}{ii}     \\  
 7 & G100 &  0.1265 &  13.6$\pm$0.7 &  15.1$\pm$0.6  &  2.4$\pm$0.7 &  72.9$\pm$1.2 &  23.0$\pm$1.1 &  43.5$\pm$2.4 & \ion{H}{ii}, (AGN)\\
 8 & G100 &  0.0576 &   3.8$\pm$1.4 &   9.3$\pm$1.2  &  1.6$\pm$0.8 &  21.3$\pm$1.1 &   1.6$\pm$1.0 &   5.3$\pm$1.3 & \ion{H}{ii},(T)\,?\\
 9 & G100 &  0.1256 &   3.8$\pm$0.7 &   0.1$\pm$0.8  &  0.3$\pm$0.6 &  25.7$\pm$1.0 &  10.9$\pm$0.9 &   5.8$\pm$1.7 & \ion{H}{ii}     \\
10 & G100 &  0.0690 &   2.4$\pm$1.0 &   0.0$\pm$0.7  &  2.3$\pm$0.7 &  29.6$\pm$1.0 &  12.9$\pm$1.0 &   2.1$\pm$1.1 & \ion{H}{ii}     \\
10 & G200 &  0.0690 &   3.4$\pm$0.3 &   2.6$\pm$0.3  &  0.5$\pm$0.4 &  33.0$\pm$0.4 &  12.5$\pm$0.4 &   3.5$\pm$0.5 & \ion{H}{ii},(T)\,? \\  
11 & G200 &  0.1824 &   0.8$\pm$0.3 &   0.4$\pm$0.3  &  0.6$\pm$0.3 &  33.1$\pm$0.6 &  12.8$\pm$0.6 &   8.2$\pm$0.5 & \ion{H}{ii}     \\
12 & G200 &  0.0576 &   0.1$\pm$0.3 &   0.1$\pm$0.4  &  2.4$\pm$0.3 &   7.0$\pm$0.4 &   1.4$\pm$0.4 &   4.4$\pm$0.5 & AGN,(\ion{H}{ii})\,?\\
13 & G100 &  0.0892 &   0.8$\pm$0.6 &  12.5$\pm$0.5  &  0.0$\pm$0.4 &   2.1$\pm$0.4 &   0.6$\pm$0.4 &   1.8$\pm$0.5 & \ion{H}{ii},(AGN)\,?\\
14 & G100 &  0.1698 &  10.0$\pm$0.6 &  18.7$\pm$0.7  &  1.8$\pm$1.0 &  65.9$\pm$1.7 &  19.6$\pm$1.4 &  22.8$\pm$1.5 & \ion{H}{ii}, (T)\\
15 & G100 &  0.0655 &  10.2$\pm$1.4 &  28.2$\pm$1.2  &  1.0$\pm$1.2 &  44.6$\pm$1.4 &   3.9$\pm$1.3 &   9.3$\pm$2.1 & \ion{H}{ii}     \\
15 & G200 &  0.0659 &   4.6$\pm$0.8 &  30.3$\pm$0.8  &  0.1$\pm$0.6 &  48.7$\pm$0.7 &   2.1$\pm$0.7 &   6.6$\pm$1.2 & \ion{H}{ii}     \\  
16 & G100 &  0.0680 &   3.3$\pm$1.4 &  16.0$\pm$1.2  &  1.3$\pm$0.9 &  15.4$\pm$1.4 &   0.0$\pm$1.3 &   0.6$\pm$1.9 & \ion{H}{ii},(AGN)\,?\\
17 & G200 &  0.1828 &   8.6$\pm$0.9 &  19.9$\pm$1.0  &  1.8$\pm$1.1 & 125.7$\pm$1.5 &  42.2$\pm$1.5 &  22.4$\pm$1.9 & \ion{H}{ii}     \\
18 & G200 &  0.1147 &   2.4$\pm$0.6 &   4.4$\pm$0.6  &  1.5$\pm$0.6 &  48.9$\pm$0.6 &  14.8$\pm$0.6 &  21.2$\pm$1.3 & \ion{H}{ii},(AGN)\\
19 & G200 &  0.0278 & 229.8$\pm$4.1 &1813.8$\pm$2.6  &  0.7$\pm$2.2 &1066.7$\pm$2.0 &  10.2$\pm$1.9 &  12.1$\pm$2.6 & \ion{H}{ii}     \\
20 & G200 &  0.2051 &   5.3$\pm$0.4 &   4.4$\pm$0.4  &  0.1$\pm$0.7 &  52.3$\pm$0.8 &  18.9$\pm$0.8 &  25.3$\pm$1.0 & \ion{H}{ii},(AGN)\\
21 & G100 &  0.1002 &   4.6$\pm$0.7 &   1.9$\pm$0.8  &  0.2$\pm$0.7 &  29.1$\pm$0.8 &   7.6$\pm$0.8 &   6.0$\pm$1.1 & \ion{H}{ii}     \\
22 & G200 &  0.1871 &   8.4$\pm$0.3 &  13.6$\pm$0.4  &  1.6$\pm$0.6 &  78.3$\pm$0.8 &  20.8$\pm$0.8 &  27.2$\pm$0.9 & \ion{H}{ii}     \\
23 & G200 &  0.2056 &   7.2$\pm$0.7 &  33.9$\pm$0.7  &  0.1$\pm$1.4 &  76.0$\pm$1.7 &  13.6$\pm$1.6 &  22.8$\pm$1.7 & \ion{H}{ii}\\
\bottomrule
\end{tabular}
\caption{
\label{comp_gal}
Spectroscopic data for the set of six comparison galaxies. 
The grism used, the measured equivalent widths (in \AA{}) of the major 
emission lines, the results of the spectral classification (S: Seyfert, 
L: LINER, T: transition type), and the classification type (u: unambiguous, 
a: ambiguous) are given.
For comparison, spectral types from the literature are given
(references: (1) Ho et al. \cite{Ho97}, (2) Greenhill et al. \cite{Gre97}).
}
\begin{tabular}{crrrrrrrlll}
\toprule
galaxy  &grism &  H$\,\beta$\,\,\,\,  & [\ion{O}{iii}]\,\, & [\ion{O}{i}]\,\,\,\, & H$\,\alpha$\,\,\,\,\, & [\ion{N}{ii}]\,\,\,\, 
& [\ion{S}{ii}]\quad\,\,\,\, &  class & class. & class\\
name    &      &  4861\,\AA{}         & 5007\,\AA{}        & 6300\,\AA{}          & 6565\,\AA{}           &  6584\,\AA{}
& 6716,6731\,\AA{}           & (here) & type   & (lit)\\
\midrule
NGC3031 & G100 & 0.0$\pm$1.1 &  11.6$\pm$0.5  &  5.3$\pm$0.2 &  11.8$\pm$0.5 &  15.4$\pm$0.4 &   5.5$\pm$0.2 & S     & u & S$^1$\\
NGC3031 & G200 & 0.1$\pm$1.7 &   8.9$\pm$0.4  &  5.4$\pm$0.2 &  19.8$\pm$0.2 &  20.3$\pm$0.2 &   5.1$\pm$0.2 & S     & u & S$^1$\\
NGC5678 & G200 & 0.1$\pm$0.3 &   0.1$\pm$0.2  &  0.3$\pm$0.2 &   2.2$\pm$0.2 &   2.4$\pm$0.1 &   0.9$\pm$0.2 & T     & u & T$^1$\\
NGC6323 & G100 & 0.0$\pm$1.9 &  41.5$\pm$2.3  &  1.1$\pm$0.5 &  14.0$\pm$1.0 &  19.8$\pm$1.0 &   7.0$\pm$1.1 & S     & u & S$^2$\\
NGC6323 & G200 & 1.9$\pm$1.2 &  31.6$\pm$0.6  &  1.4$\pm$0.3 &  22.0$\pm$0.4 &  21.2$\pm$0.4 &   6.1$\pm$0.4 & S     & u & S$^2$\\
NGC6500 & G200 & 2.7$\pm$1.7 &  14.0$\pm$0.4  & 10.5$\pm$0.3 &  45.8$\pm$0.2 &  29.9$\pm$0.2 &  36.4$\pm$0.3 & L,(S) & a & L$^1$\\
NGC7177 & G200 & 0.1$\pm$0.4 &   0.1$\pm$0.3  &  0.1$\pm$0.2 &   5.2$\pm$0.2 &   3.0$\pm$0.1 &   5.1$\pm$0.2 & T?    & a & T$^1$\\
NGC7743 & G200 & 0.1$\pm$0.3 &   5.1$\pm$0.3  &  1.1$\pm$0.2 &   9.2$\pm$0.2 &  13.3$\pm$0.2 &   7.4$\pm$0.2 & T,(S) & u & S$^1$\\
\bottomrule
\end{tabular}
\end{table*}

It is a common practice to discriminate between AGNs and massive stars
by means of line-ratio diagrams
(Baldwin et al. \cite{Bal81}; Vielleux \& Osterbrock \cite{Vie87};
Ho et al. \cite{Ho97}). We use the diagnostic line-ratios recommended
by Vielleux \& Osterbrock that are based on
the measurements of the lines mentioned above.
Since the wavelength separation of the considered lines are small,
these line-ratios are relatively insensitive to reddening
(cf. Dessauges-Zavadsky et al. \cite{Des00}).
Line-ratios are expressed by the ratios of the corresponding EWs.
Figure\,\ref{diagnostics} shows these  diagnostic diagrams
for all galaxies. Each object was individually classified
on each of the three diagrams, and then the consistence of the
three classifications was evaluated.
We consider three spectral classes:
\ion{H}{ii} galaxies to the left of the demarcation curves,
AGNs to the right, and transition objects
near the boundary (i.e., the error bars cross the demarcation),
In addition, we define two classification types: unambiguous (u),
i.\,e. all relevant data are consistent with one spectral class, or
ambiguous (a), i.e. the object is classified both as
AGN and \ion{H}{ii} galaxy on different diagrams. The results
are given in  Table\,\ref{spectra}.
For ambiguous classifications the two most likely classes are given.
There is good agreement between the spectral classes derived for the six
comparison galaxies and the classification from the literature
(Table\,\ref{comp_gal}).

The results can be summarised as follows:
for most of the NELGs the line-ratios correspond to \ion{H}{ii} region
spectra. None of the NELGs is unambiguously classified as an AGN.
Only for one object is more than one diagram compatible with
an AGN spectrum, but the signal-to-noise is relatively low in
this case. A substantial fraction of the sample (40\%) has an
ambiguous classification, and a similar fraction is
located near the \ion{H}{ii}-AGN border.
It must be emphasised that the line-ratios are derived from
integrated spectra. At a redshift of $z=0.1$, a slit width of $1''$
covers about 2\,kpc. In the direction of the long-slit,
the spectrum integrates over the whole galaxy.
For many of the NELGs rotation curves are seen in H$\alpha$
indicating the presence of extended emission regions like
circum-nuclear rings or spiral arms.
If an LLAGN is present, it can thus be masked by the emission
from \ion{H}{ii} regions connected to intense
star formation (Storchi-Bergmann et al. \cite{Sto96};
Pastoriza et al. \cite{Pas99}).
A trend of decreasing line ratios with increasing effective aperture,
and therewith with increasing $z$, is expected if AGNs
substantially contribute to the spectra (Storchi-Bergmann \cite{Sto91}).
Such a trend is not indicated in our data.
We cannot exclude that there are LLAGNs in at least some
of the NELGs. However, the integrated spectra are not
dominated by AGNs.


\begin{figure*}[hpbt]
 \begin{tabbing}
 \resizebox{8.0cm}{7.0cm}{\includegraphics{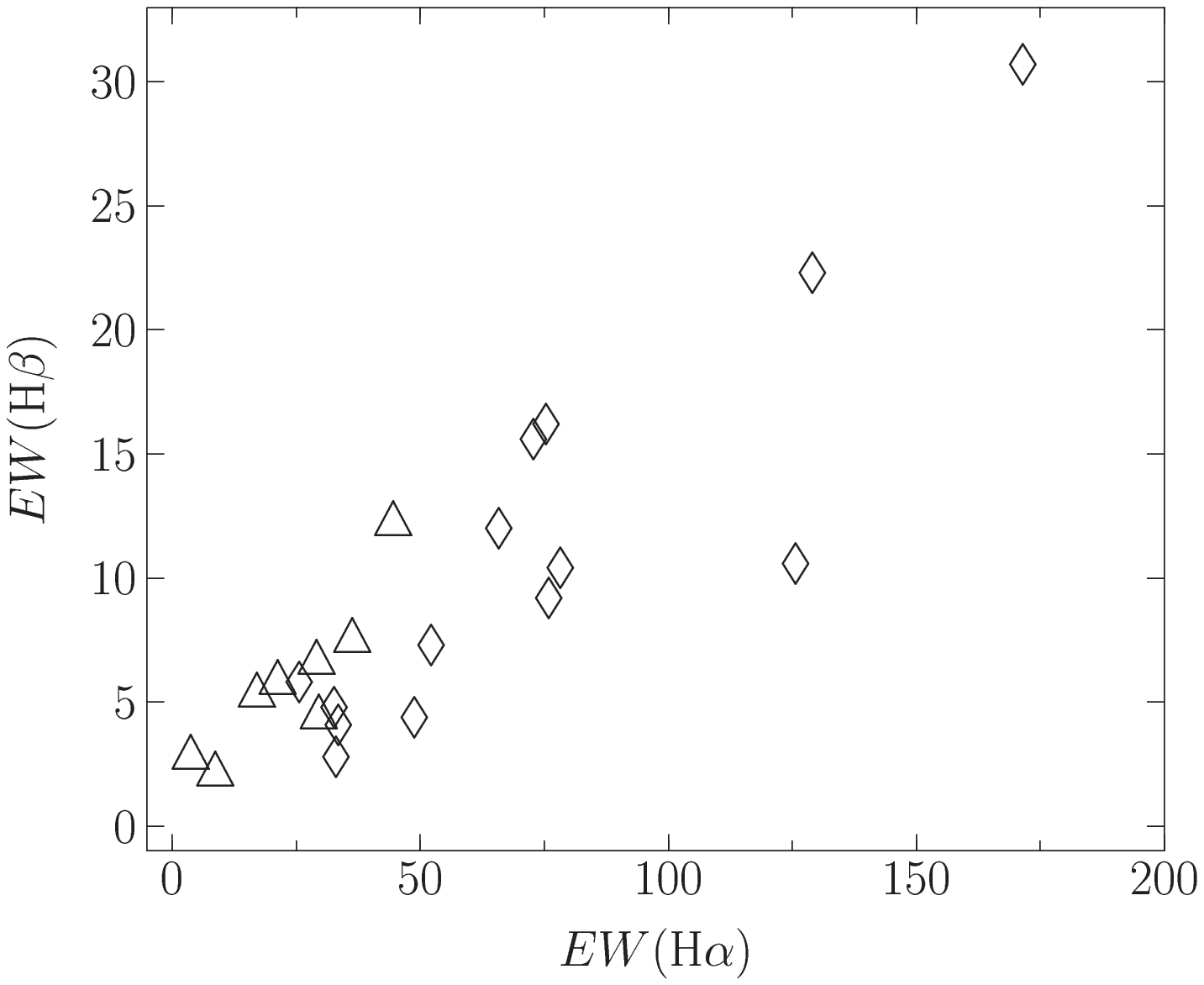}} \hfill \=
 \resizebox{8.0cm}{7.0cm}{\includegraphics{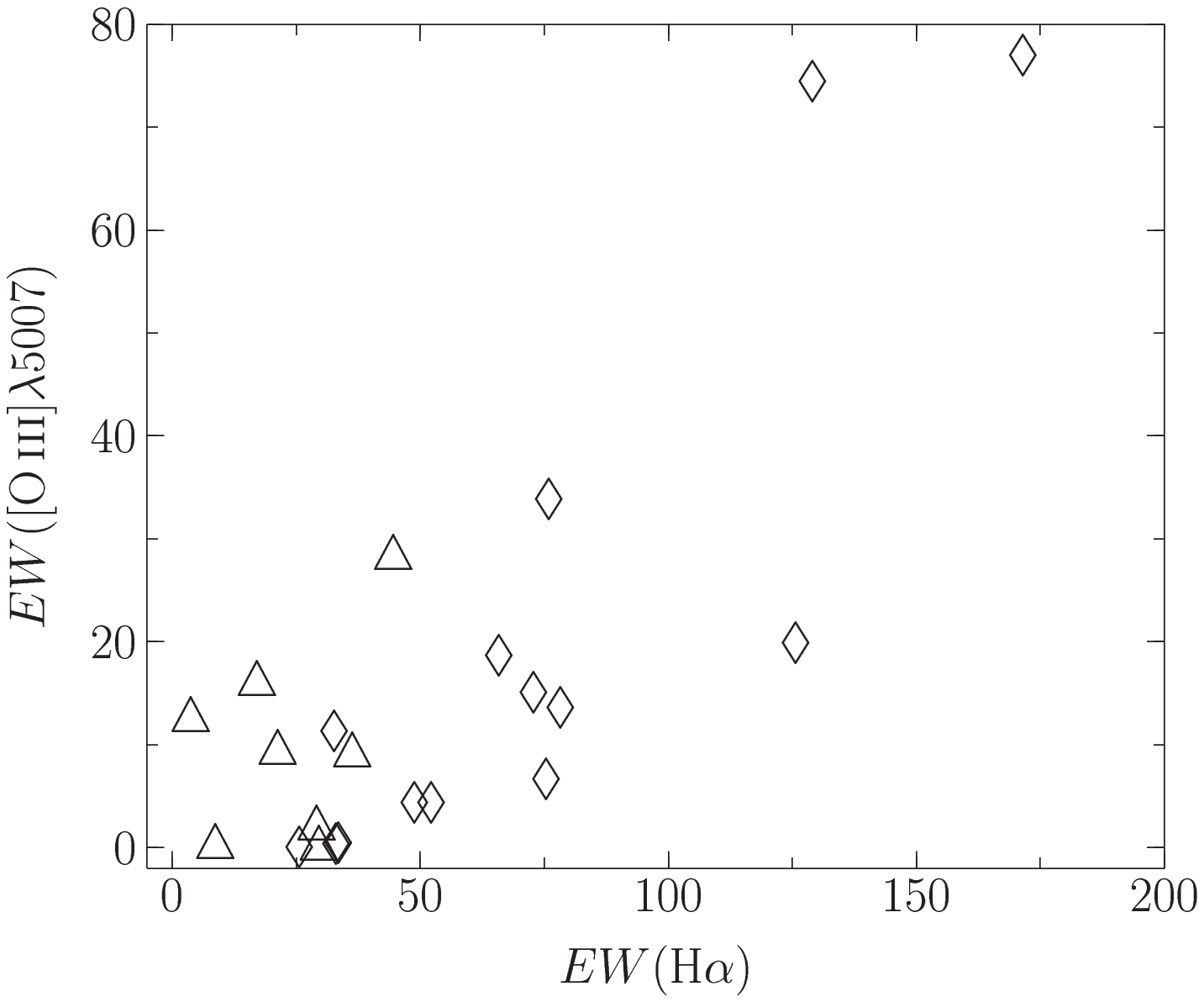}}\= \\
 \resizebox{8.0cm}{7.0cm}{\includegraphics{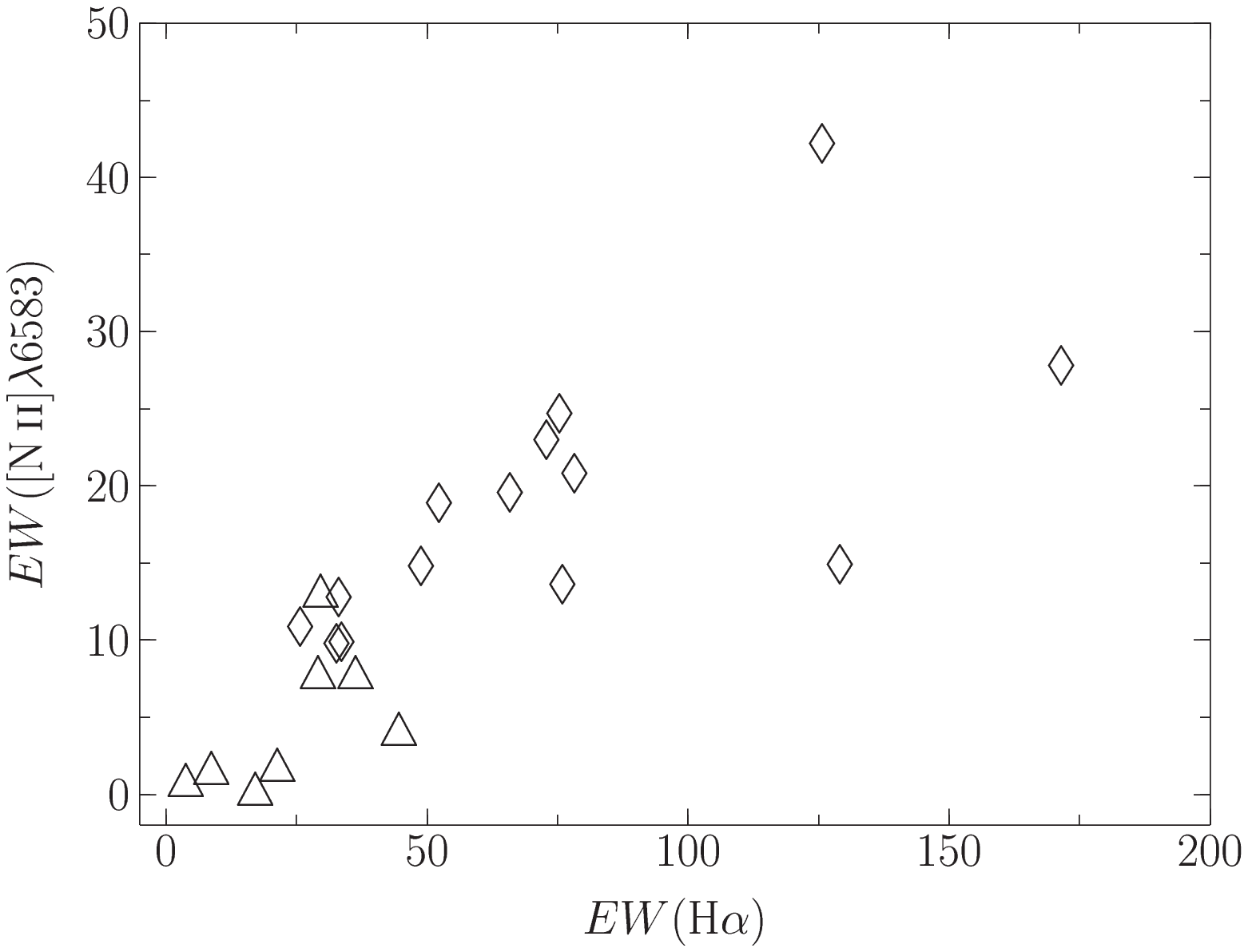}}\>
 \resizebox{8.0cm}{7.0cm}{\includegraphics{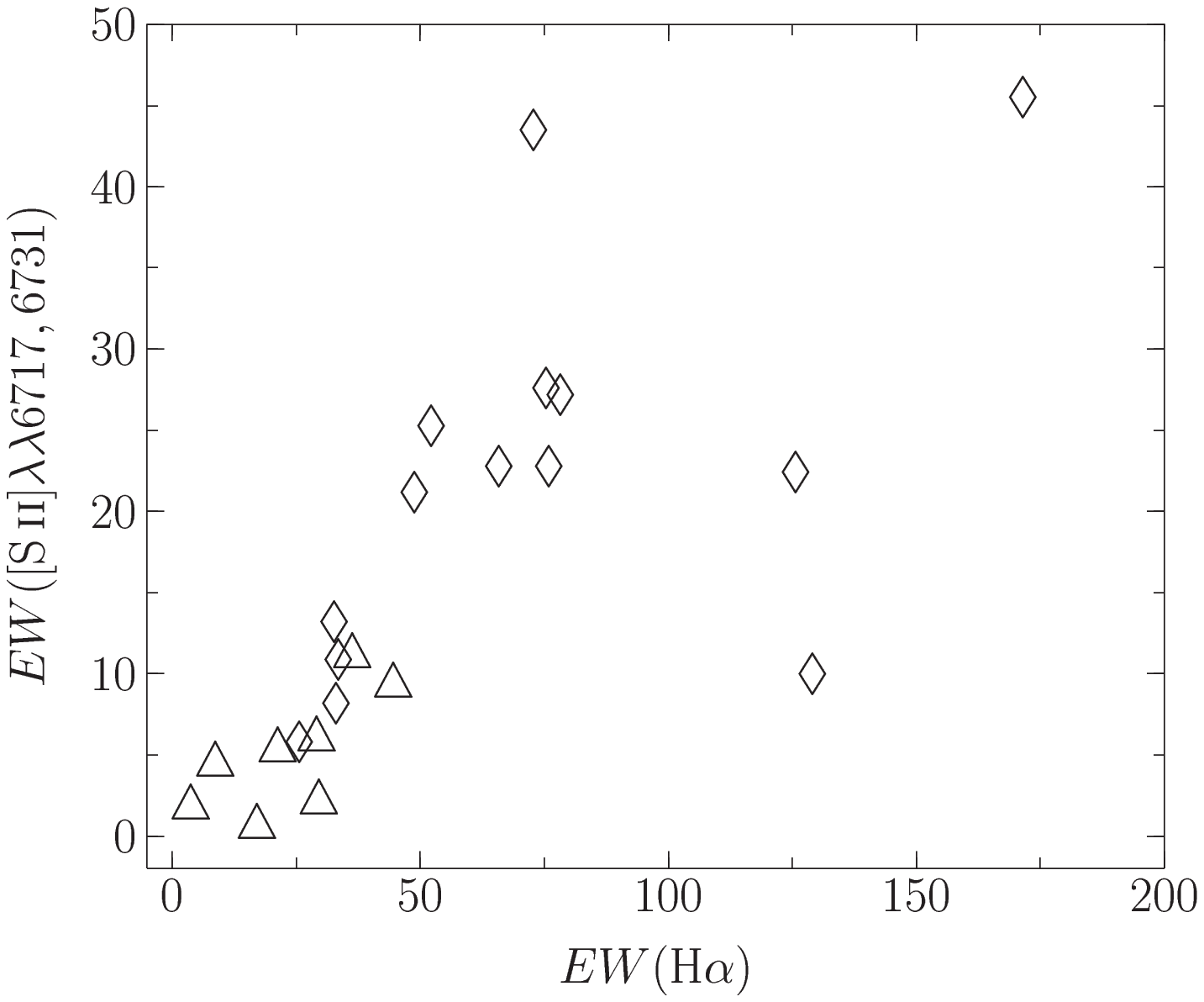}}\> \\
 \resizebox{8.0cm}{7.0cm}{\includegraphics{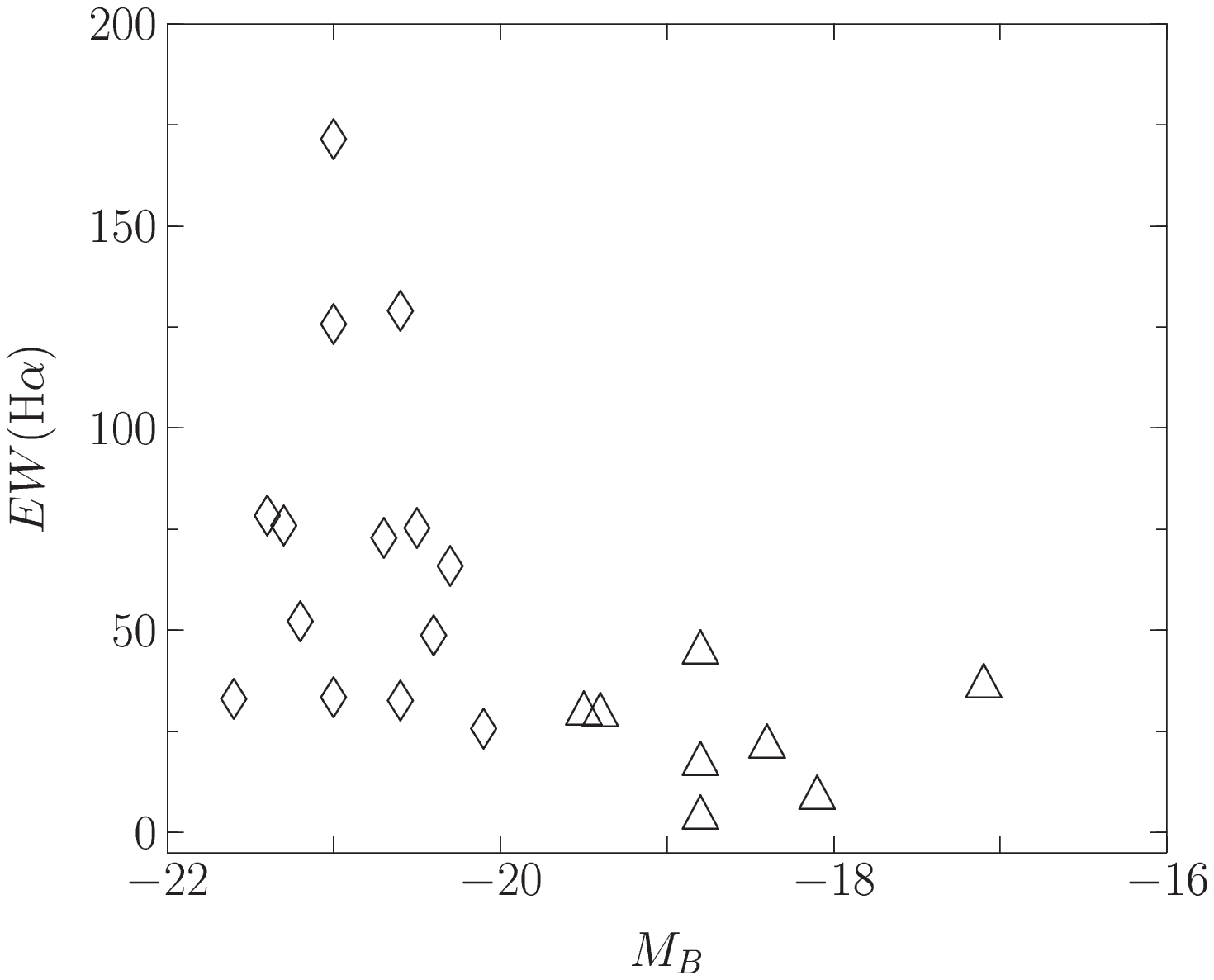}}\>
 \resizebox{8.0cm}{7.0cm}{\includegraphics{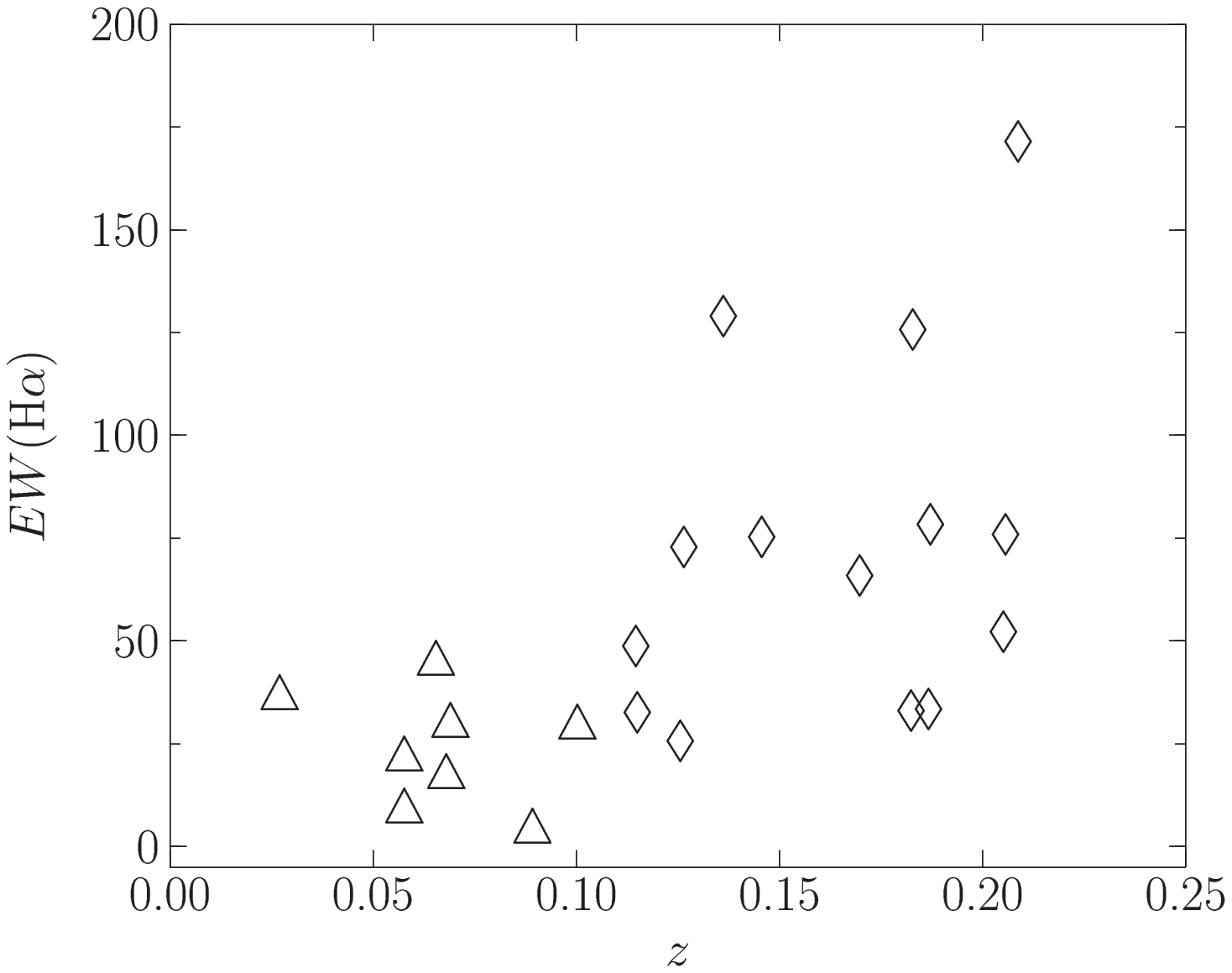}}\> \\
 \end{tabbing}
\caption{
Correlations with $EW(\mbox{H}\alpha$).
Symbols as in Fig.\,\ref{class}.
\label{corr}
}
\end{figure*}


%
\section{Further properties of the NELG sample}

Some trends with $EW(\mbox{H}\alpha)$ are shown in
Fig.\,\ref{corr} (where NELG\,19 with its exceptionally
high EWs was excluded for the sake of clearness). There is
a strong correlation between $EW(\mbox{H}\beta$) and $EW(\mbox{H}\alpha$).
The slope is similar to that of nearby galaxies and implies a
mean extinction $A_{V} \approx 1$\,mag (cf. Kennicutt \cite{Ken92}).
Further, there are loose correlations with $EW(\mbox{H}\alpha$) for the
EWs of the lines
[\ion{O}{iii}]$\lambda$5007,
[\ion{N}{ii}]$\lambda$6583,
and [\ion{S}{ii}]$\lambda\lambda$6717,6731.
Such trends are known from samples of nearby galaxies
(e.g., Tresse et al.\cite{Tre99};
Sodr\'e \& Stasi\'nska \cite{Sod99}).
The large dispersions are likely due to
the variation in the mean nebular excitation in the
galaxies (Kennicutt \cite{Ken92}).

Our sample is magnitude-limited and the absolute magnitudes are
therefore strongly correlated with redshifts. Further,
due to the constraint of star-like images, the
characteristic scale-length of the dominant component
is of the order of a few arcseconds or less.
The sample thus comprises roughly two classes of objects:
dwarfs and sub-$L^{\ast}$ galaxies at $z<0.1$ and $\sim L^{\ast}$
galaxies at $0.1 \la z \la 0.2$. These two subclasses are
marked by different symbols in all relevant figures.
The most significant difference between the sub-$L^{\ast}$
and the $\sim L^{\ast}$ galaxies are the mean EWs
(Fig.\,\ref{corr}, bottom):
NELGs with fainter absolute magnitudes show a clear tendency to
have smaller $EW(\mbox{H}\alpha$). This trend is opposite to
what is found in a representative sample of galaxies in the
local universe (Tresse et al. \cite{Tre99}).
If the trend in Fig.\,\ref{corr} is real, it is probably
related to the selection effects of the VPM search.
However, a plausible explanation for this trend
has not yet been found.

Most of the NELGs are blue (Table\,\ref{foto}). Using the
$K$ corrections for Sbc galaxies (Coleman et al. \cite{Col80}),
the corrected sample-averaged colour indices are $\langle U-B
\rangle = -0.66\pm0.37$ and $\langle B-V \rangle = 0.48\pm0.25$.
(The $U-B$ colour is probably slightly underestimated due to a bias
towards brighter magnitudes at the faint end of the $U$
measurements.) The blue colours are likely a
selection effect: the object selection is based on the
morphological classification done on the deepest red plate and on the
variability measured on $B$ plates (Paper~1). Galaxies that are
compact on the red plate but more extended, and therewith brighter
on the blue plates, have therefore a good
chance to be selected as (nearly) star-like and variable objects.
The $M_B$ distribution of the NELGs is similar to that
of galaxies selected for their compact nuclei (Sarajedini
et al. \cite{Sar99}).

\input 2130f07

\begin{figure*}[hpbt]
\vspace{1.0cm}
\resizebox{16.04cm}{21.2cm}{\includegraphics{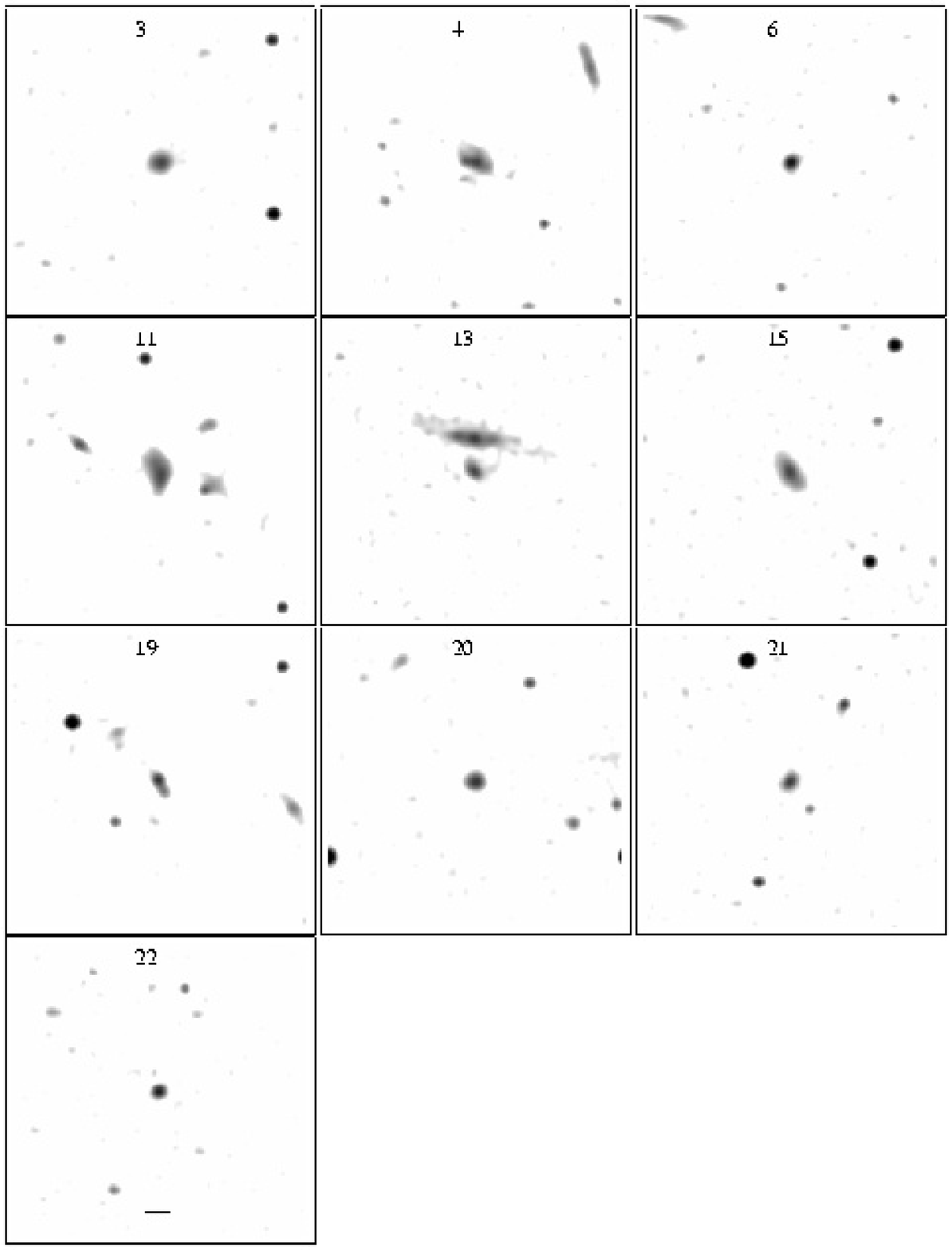}}
\caption{
Logarithmic gray scale presentation of 10 NELGs (in the middle
of each panel) in the $B$ band after Lucy-Richardson deconvolution.
The image size is 1\arcmin$\times$1\arcmin, the horizontal
line indicates 20\,kpc at the distance of the galaxy. The
panel numbers correspond to the running numbers of the
NELGs from Table\,\ref{foto}.
N is up, E is right.
\label{images}}
\end{figure*}


The $EW(\mbox{H}\alpha)$ vs. $B-V$ diagram for the NELGs of the
present study  (Fig.\,\ref{Ha_B_V}) resembles the one for the
starburst galaxies discussed by Moy et al. (\cite{Moy01}).
According to these authors, the most attractive explanation
for this diagram is provided by galaxy models with continuous star
formation + starbursts where the  ages and metallicities
of the underlying population are similar to those in normal
spirals.

Direct 100\,s exposures were obtained in the $V$ band for all 23 NELGs.
There is little information
about morphological details for most of the galaxies.
Deeper images are
available for the 10 NELGs from the photometric series
where six 180\,s $B$ band exposures for each object were co-added.
Since we are interested in the extended, non-star-like
structure components, the images were PSF-deconvolved
applying the Lucy-Richardson algorithm.
The resulting images (Fig.\,\ref{images}) show that all
NELGs have dominant, more or less compact components.
In addition, a more extended fainter component is indicated in many cases.
Structure details like spiral arms, bars, or rings cannot be
recognised. Some NELGs (4, 11, 19) clearly show asymmetric
light distributions,
perhaps indicating morphological distortions. Many of the
NELGs have several close-by neighbour galaxies (in projection).
For instance, NELG\,11 seems to be a highly disturbed galaxy
in the centre of a group, and the dwarf NELG\,13 seems to
have a light-bridge toward a nearly edge-on giant spiral at a projected
distance of less than 20\,kpc. A further interesting object is NELG\,19
which is probably a cometary blue compact dwarf galaxy with EWs
similar to Mkn\,71 (for the latter, see Kennicutt \cite{Ken92};
Noeske et al. \cite{Noe00}).

The VPM NELGs have properties characteristic of starbursts and are thus
expected to be strong emitters in the far infrared (FIR).
There are no entries in the NED$^1$
\footnotetext[1]{
The NASA/IPAC Extragalactic Database (NED) is operated by the Jet
Propulsion Laboratory, California Institute of Technology, under
contract with the National Aeronautics and Space Administration.}
at the positions of the NELGs, and there are no
IRAS counterparts. The latter is simply explained by the IRAS
detection limits: galaxies with $B\approx19$
must have $\log\,L_{\rm FIR}/L_{\rm B} > 1.7$
to be found in the IRAS catalogues. Such a strong
FIR excess is characteristic of only a few ultra-luminous IR
galaxies (Sanders \& Mirabel \cite{San96}) which have
much stronger internal extinction in the optical than is
indicated for the NELGs (Fig.\,\ref{corr}).

\section{Conclusions}

We have studied the sample of NELGs from the VPM survey in the
M\,92  field. These objects have been selected  as QSO candidates
because of their high variability indices (Paper~1) and have
been classified later as NELGs (Paper~2). However, it was not clear
from the previous data to what extent the variability and the
spectral properties of the NELGs are related to AGNs. In the
present paper, we re-investigated the variability and analysed the
emission line-ratios, as well as the photometric and morphological
properties of the NELGs. The main conclusions are the following.
\begin{itemize}
\item{The variability indices reported in Paper~2
      primarily reflect increased measurement errors due to the
      resolved image profiles and do not provide evidence for AGNs.
      The measurement of variability for resolved objects requires
      techniques other than Gaussian profile fitting, even when the
      deviations from the stellar profiles are small.
      An unambiguous separation between stellar and nonstellar
      objects is crucial for the VPM survey.
      }
\item{The diagnostic line-ratio diagrams are best explained
      by \ion{H}{ii} region-like spectra. None of the NELGs is
      unambiguously classified as an AGN. At least for some of the
      NELGs, the existence of LLAGNs  cannot be excluded.
      However, if present, AGNs do not dominate the integrated spectra.
      }
\item{The VPM NELGs are compact, blue galaxies.
      Most of them are related to starbursts.
      The sample consists of a range of various types,
      as is known from other samples of local starburst galaxies
      (e.\,g., Coziol et al. \cite{Coz98}).
      }
\end{itemize}

An important result of this work is the substantial
improvement of both the photometric accuracy and
the star-galaxy separation for the objects from the VPM survey
in the M\,92 field. 
This enabled us to identify additional VPM-QSO candidates
with high or medium priority in our sample. Their
spectroscopic observations, the newly detected QSOs 
and the properties of the enlarged QSO
sample will be the subject of Paper~4 of this series.

%
\begin{acknowledgements}
%
This research is based on observations made with the 2.2m telescope
of the German-Spanish Astronomical Centre, Calar Alto, Spain.
We acknowledge financial support from the
\emph{Deut\-sche For\-schungs\-ge\-mein\-schaft} under 
grants Me1350/8 and /10.
This research has made use of the NASA/IPAC Extragalactic
Database (NED) which is operated by the Jet
Propulsion Laboratory, California Institute of Technology, under
contract with the National Aeronautics and Space Administration.

\end{acknowledgements}


\end{document}

%% file: 2130f01
\begin{figure}
\beginpicture
\unitlength2mm
\setcoordinatesystem units <15.5mm,100mm>
\setplotarea x from -0.5 to 4.2, y from -0.9 to -0.3
\axis top 
      ticks in long unlabeled from 0 to 4 by 1
              short unlabeled from 0 to 4 by 0.5
/
\axis bottom label {$\log\,\tau\,\mbox{(days)}$}
      ticks in long numbered  from 0 to 4 by 1
              short unlabeled from 0 to 4 by 0.5
/
\axis left label
{\begin{sideways}$\log\,SF$ \end{sideways}}
      ticks in long numbered  from -0.8 to -0.4 by 0.2
              short unlabeled from -0.8 to -0.4 by 0.1
/              
\axis right 
      ticks in long unlabeled from -0.8 to -0.4 by 0.2
              short unlabeled from -0.8 to -0.4 by 0.1
/
\multiput {$\triangle$} at 
-0.30 -0.485
 1.01 -0.485
 1.74 -0.491
 2.06 -0.472
 2.48 -0.429
 2.80 -0.422
 3.10 -0.477
 3.40 -0.428
 3.70 -0.421
 3.85 -0.446
 4.00 -0.542
/%
{\normalsize
\multiput {$\bigstar$} at
-0.30   -0.891
1.01   -0.841
1.74   -0.851
2.06   -0.799
2.48   -0.771
2.80   -0.711
3.10   -0.646
3.40   -0.619
3.70   -0.614
3.85   -0.640
4.00   -0.711
/%
}%
\setsolid
\plot
-0.30 -0.485
 1.01 -0.485
 1.74 -0.491
 2.06 -0.472
 2.48 -0.429
 2.80 -0.422
 3.10 -0.477
 3.40 -0.428
 3.70 -0.421
 3.85 -0.446
 4.00 -0.542
/%
\setsolid
\plot
-0.30   -0.891
1.01   -0.841
1.74   -0.851
2.06   -0.799
2.48   -0.771
2.80   -0.711
3.10   -0.646
3.40   -0.619
3.70   -0.614
3.85   -0.640
4.00   -0.711
/%
\put {NELGs} at 1.2 -0.42
\put {QSOs} at 1.2 -0.8
\endpicture
\caption{\label{sf}
Sample averaged structure function $SF$ as a function of the time-lag $\tau$
in days for the NELGs (open triangles and upper 
polygon) and the QSOs (asterisks and lower polygon), respectively, from
the VPM survey in the M\,92 field.}
\end{figure}

%% file: 2130f02
\begin{figure}[htbp]
\unitlength1mm
\beginpicture
\setcoordinatesystem units <1.5mm,1.5mm> point at 55 -5
\setplotarea x from 0 to 40, y from -5 to 40
\axis bottom
      label {$I_\sigma^{\rm old}$}
      ticks in long numbered from 0 to 40 by 10
               short unlabeled from 0 to 40 by 5
/
\axis left
      label {\begin{sideways}$I_\sigma^{\rm new}$\end{sideways}}
      ticks in long numbered from 0 to 40 by 10
               short unlabeled from -5 to 40 by 5
/
\axis right
      ticks in long unlabeled from 0 to 40 by 10
               short unlabeled from -5 to 40 by 5
/
\axis top
      ticks in long unlabeled from 0 to 40 by 10
               short unlabeled from 0 to 35 by 5
/
\put {$\bullet$} at 16.61387  35.5184    
\put {$\nearrow$} [tr] at 15.61387  34.5184
\put {\scriptsize behind foreground galaxy} [t] at 17.5 31
\put {$\bullet$} at 19.91420   4.3548    
\put {$\bullet$} at 32.47946   5.7015    
\put {$\bullet$} at  9.18736  -0.7221    
\put {$\bullet$} at  8.96030   0.6935    
\put {$\bullet$} at 10.61244   3.3249    
\put {$\bullet$} at 11.68180   4.2025    
\put {$\bullet$} at 10.82710   2.8516    
\put {$\bullet$} at 11.85133   4.2317    
\put {$\bullet$} at  3.65272   1.4579    
\put {$\bullet$} at  6.26015   2.1133    
\put {$\bullet$} at  9.76339   2.2755    
\put {$\bullet$} at  6.41008   0.7498    
\put {$\bullet$} at  9.54378   2.8566    
\put {$\bullet$} at  6.16013  -0.0340    
\put {$\bullet$} at 13.66801   3.1080    
\put {$\bullet$} at  3.67757   0.2896    
\put {$\bullet$} at  5.53530   1.4389    
\put {$\bullet$} at  6.74909  -2.0170    
\put {$\bullet$} at 10.37503   2.9673    
\put {$\bullet$} at  5.82375   1.8897    
\put {$\bullet$} at  7.57117   0.0712    
\put {$\bullet$} at 14.51876   2.6568    
\put {$\bullet$} at  5.19894   2.8743    
\put {$\bullet$} at  6.78267  -1.0004    
\put {$\bullet$} at  5.30098   0.5589    
\put {$\bullet$} at  7.92899   0.8682    
\endpicture

\vspace{0.3cm}
\beginpicture
\setcoordinatesystem units <10mm,10mm> point at -3 -3
\setplotarea x from -3 to 3, y from -3 to 3
\axis bottom
      label {$I_\Delta^{\rm old}$}
      ticks in long numbered from -2 to 2 by 2
               short unlabeled from -3 to 3 by 1
/
\axis left
      label {\begin{sideways}$I_\Delta^{\rm new}$\end{sideways}}
      ticks in long numbered from -2 to 2 by 2
               short unlabeled from -3 to 3 by 1
/
\axis right
      ticks in long unlabeled from -2 to 2 by 2
               short unlabeled from -3 to 3 by 1
/
\axis top
      ticks in long unlabeled from -2 to 2 by 2
               short unlabeled from -3 to 3 by 1
/
\put {$\bullet$} at  0.37389   1.0029  
\put {$\bullet$} at  1.99578   0.4780  
\put {$\bullet$} at  0.45629  -2.0450  
\put {$\bullet$} at  1.75568  -0.2914  
\put {$\bullet$} at  0.37163   1.6616  
\put {$\bullet$} at -0.22734  -1.7492  
\put {$\bullet$} at  1.54563  -0.1416  
\put {$\bullet$} at  0.44666  -0.8578  
\put {$\bullet$} at -1.25878   0.4092  
\put {$\bullet$} at  0.55370   0.6455  
\put {$\bullet$} at  2.04540  -0.2077  
\put {$\bullet$} at  0.51529   2.5740  
\put {$\bullet$} at  1.30566   0.1669  
\put {$\bullet$} at -0.40022  -1.2104  
\put {$\bullet$} at  0.91473  -0.2162  
\put {$\bullet$} at -2.00326   0.1089  
\put {$\bullet$} at  1.66838   0.4258  
\put {$\bullet$} at  0.33212   0.6916  
\put {$\bullet$} at -1.79156  -0.0802  
\put {$\bullet$} at  1.39075   0.7680  
\put {$\bullet$} at -1.49308   1.7075  
\put {$\bullet$} at  1.38113   1.2128  
\put {$\bullet$} at -2.41663  -0.1967  
\put {$\bullet$} at -0.57671  -1.3833  
\put {$\bullet$} at -0.18490   0.2581  
\put {$\bullet$} at  0.72024  -0.6152  
\put {$\bullet$} at  1.31886  -0.0429  
\endpicture
\caption{\label{old_new} 
New and old variability indices for the NELGs for overall variability
(top) and long-term variability (bottom).
}
\end{figure}

%% file: 2130f04
\begin{figure}[h]
\unitlength1mm
\beginpicture
\setcoordinatesystem units <42mm,37mm> point at 0 0
\setplotarea x from 0.4 to 2.2, y from 0 to 1.65
\axis bottom
      label {$\log\ I_{\rm nonstellar}^{\rm new}$}
      ticks in long numbered from 0.5 to 2 by 0.5
               short unlabeled from 0.5 to 2.2 by 0.1
/
\axis left
      label {\begin{sideways}$\log\ I_{\sigma}^{\rm new}$\end{sideways}}
      ticks in long numbered from 0 to 1.5 by 0.5
               short unlabeled from 0.1 to 1.6 by 0.1
/
\axis top
      ticks in long unlabeled from 0.5 to 2 by 0.5
               short unlabeled from 0.5 to 2.2 by 0.1
/
\axis right
      ticks in long unlabeled from 0.5 to 1.5 by 0.5
               short unlabeled from 0.1 to 1.6 by 0.1
/
\input 2130f04.dat
\endpicture
\caption{\label{norm_diff}
Overall variability index $I_\sigma^{\rm new}$ versus image profile
index $I_{\rm nonstellar}^{\rm new}$ from the revised photometry 
for the NELGs with significant variability, i.\,e. $I_\sigma^{\rm new} > 1$.
(Symbols as in Fig.\,\ref{class}.)
}
\end{figure}

%% file: 2130f07
\begin{figure}
\beginpicture
\unitlength2mm
\setcoordinatesystem units <50mm,20mm>
\setplotarea x from -0.3 to 1.25, y from 0 to 3.5
\axis top 
      ticks in long unlabeled from 0 to 1. by 0.5
              short unlabeled from -0.3 to 1.2 by 0.1
/
\axis bottom label {$B-V$}
      ticks in long numbered  from 0 to 1 by 0.5
              short unlabeled from -0.3 to 1.2 by 0.1
/
\axis left label
{\begin{sideways}$\log\,EW(\mbox{H}\alpha)$ \end{sideways}}
      ticks in long numbered  from 0 to 3 by 1
              short unlabeled from 0 to 3.5 by 0.5
/              
\axis right 
      ticks in long unlabeled from 0 to 3 by 1
              short unlabeled from 0 to 3.5 by 0.5
/
\input 2130f07.dat
\setsolid
\plot
 0.15 1.56
 0.04 1.74
-0.11 2.38
-0.20 2.80
-0.18 2.88
/%
\plot
 0.65  1.36
 0.56  1.54
 0.435 2.18
 0.41  2.63
 0.47  2.71
/%
\plot
 1.05  1.13
 0.92  1.31
 0.83  1.95
 0.855 2.40
 0.92  2.48
/%
\setsolid
\plot
0.15 1.56
0.65 1.36
1.05 1.13
/%
\setsolid
\plot
0.04 1.74
0.56 1.54
0.92 1.31
/%
\setsolid
\plot
-0.11  2.38
 0.435 2.18
 0.83  1.95
/%
\setsolid
\plot
-0.20  2.80
 0.41  2.63
 0.855 2.40
/%
\setsolid
\plot
-0.18  2.88
 0.47  2.71
 0.92  2.48
/%
\setdots
\plot
0.30 1.355
0.15 1.55
/%
\plot
0.73 1.13
0.65 1.36
/%
\plot
1.07 0.885
1.05 1.13
/%
\plot
0.3  1.355
0.73 1.13
1.07 0.885
/%
\put {0 Myr} at 1.05 2.5
\put {2 Myr} at 0.96 2.32
\put {4 Myr} at 0.95 1.9
\put {6 Myr} at 1.03 1.35
\put {8 Myr} at 1.15  1.13
\put {$\tau_V^{\rm neb}=0$} at -0.13 3.0
\put {$\tau_V^{\rm neb}=2$} at 0.5 2.84
\put {$\tau_V^{\rm neb}=4$} at 0.9 2.64
\endpicture
\caption{\label{Ha_B_V}
$EW(\mbox{H}\alpha)$ versus $B-V$ for the 
NELGs (symbols as in Fig.\,\ref{class}). The solid lines show the predictions 
of model D by Moy et al. (\cite{Moy01}; their Fig.\,10) 
for continuous star formation + starbursts of different  
ages (labeled to the right of the grid) 
and extinction levels (labeled by the optical depth $\tau_V^{\rm neb}$
at the top of the grid)
for metallicity $Z=1/2\,Z_{\odot}$. 
The dotted lines show the shift of the low part of the grid 
for $Z=3/2\,Z_{\odot}$.
}
\end{figure}